\documentclass[aps,pra,twocolumn,showpacs,10pt,amsmath,footinbib]{revtex4-1}
\usepackage{graphicx}
\usepackage[usenames,dvipsnames]{xcolor}
\usepackage{braket}
\usepackage{array}

\usepackage{tikz}
\usepackage{pgfplots}

\newcommand{\pa}{\partial}

\def\gappeq{\mathrel{ \rlap{\raise.5ex\hbox{$>$}}
                      {\lower.5ex\hbox{$\sim$}} } }
\def\lappeq{\mathrel{ \rlap{\raise.5ex\hbox{$<$}}
                      {\lower.5ex\hbox{$\sim$}} } }
\begin{document}

\title{Ultra-quantum turbulence in a quenched homogeneous Bose gas}

\author{G. W. Stagg}\email{george.stagg@ncl.ac.uk}
\affiliation{Joint Quantum Centre (JQC) Durham--Newcastle, School of Mathematics and Statistics, Newcastle University, Newcastle upon Tyne, NE1 7RU, United Kingdom}
\author{N. G. Parker}
\affiliation{Joint Quantum Centre (JQC) Durham--Newcastle, School of Mathematics and Statistics, Newcastle University, Newcastle upon Tyne, NE1 7RU, United Kingdom}
\author{C. F. Barenghi}
\affiliation{Joint Quantum Centre (JQC) Durham--Newcastle, School of Mathematics and Statistics, Newcastle University, Newcastle upon Tyne, NE1 7RU, United Kingdom}

\date{\today}

\begin{abstract}
Using the classical field method, we study numerically the characteristics and decay of the
turbulent tangle of superfluid vortices which is created in the evolution of
a Bose gas from highly nonequilibrium initial conditions. By analysing the vortex line density, the 
energy spectrum and the velocity correlation function, we determine
that the turbulence resulting from this effective thermal quench lacks the
coherent structures and the Kolmogorov scaling; these properties are
typical of both
ordinary classical fluids and of superfluid helium when driven by grids
or propellers. Instead, thermal quench turbulence has properties 
akin to a random flow, more
similar to another turbulent regime called ultra-quantum turbulence which has been 
observed in superfluid helium.
\end{abstract}

\maketitle

\section{Introduction}
\begin{figure*}
\centering%
\includegraphics[width=\linewidth]{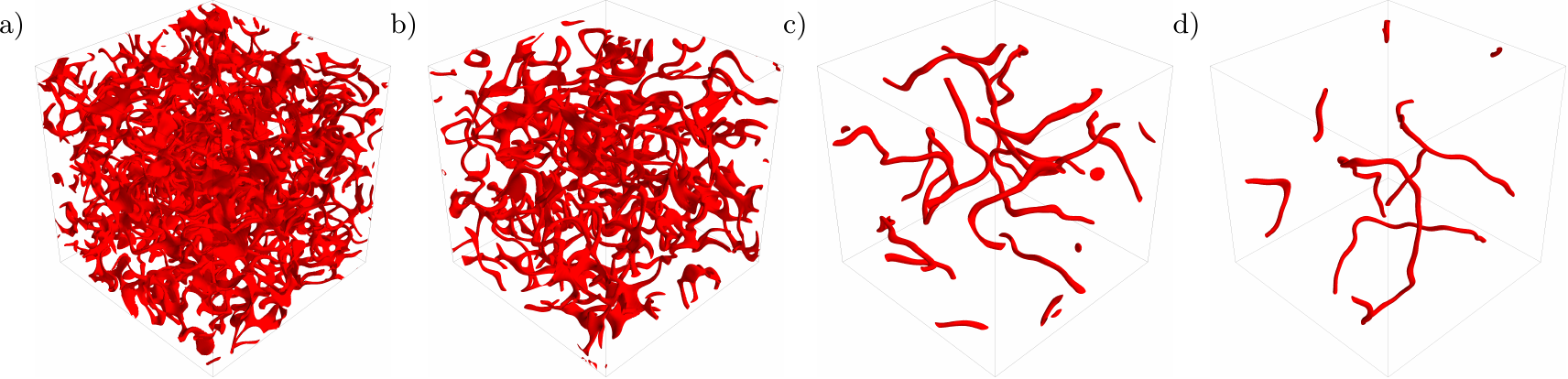}
\caption{(Color online).
Sample evolution of the turbulent vortex tangle (present in the 
condensed part of the Bose gas) as it decays to equilibrium. 
Here the condensate fraction of the gas is $\rho_0/\rho = 0.22$. 
Shown are iso-surfaces of the quasi-condensate density at 
isosurface level $0.05\braket{|\hat{\psi}|^2}$ at times 
(a) $t/\tau=0$, (b) $250$, (c) $1250$, and (d) $2500$. 
At later times (not shown) all vortex lines disappear from the system. 
Here the quasi-condensate is visualised using a cutoff of $k_{c} = 10~(2\pi/D)$.  In the Supplemental Material \cite{supp}, we provide a movie of this process. A similar figure and further description of the quasi-condensate 
filtering are shown in Ref.~\cite{Berloff2002}.}
\label{fig1}
\end{figure*}

The formation of a coherent Bose-Einstein condensate from a thermal Bose 
gas is a rich topic of ongoing research \cite{Davis2016}. 
Recent experiments on 
thermally-quenched Bose gases have observed the spontaneous formation 
of defects in the guise of vortices \cite{Weiler2008,Chomaz2015} and 
solitonic vortices \cite{Lamporesi2013}, confirming the occurrence 
of the Kibble-Zurek mechanism \cite{Kibble1976,Zurek1985} in these gases.  
A paradigm for this non-equilibrium phase transition  
is the formation of a homogeneous weakly-interacting Bose 
gas starting from highly non-equilibrium initial conditions 
\cite{Svis1,Svis2,Svis3,Svis4,Svis5,Davis,Berloff2002,Davis2,
Connaughton2005}. 
The gas is modelled as a classical matter field described by the Gross-Pitaevskii equation. This field undergoes a universal self-ordering into a quasi-condensate, i.e. a coherent superfluid component, and a non-condensed, thermal component.  Phase dislocations that become entrapped within the quasi-condensate during this quench give rise to an irregular tangle of quantized vortex lines that permeate the system, as seen in Fig. ~\ref{fig1}.  This corresponds to a state of superfluid turbulence. Over time, this turbulent state relaxes, tending towards 
the vortex-free, partially-condensed equilibrium state of the finite-temperature Bose gas.  

The aim of our work is to explore the nature of this 
state of superfluid turbulence, including the characteristics of its decay. Is it similar to turbulence in ordinary (classical) fluids?
The question is natural, because tangles of quantized vortices
created in superfluid helium by moving grids or propellers appear to obey
classical scaling laws \cite{Tabeling1998,Salort2010,Barenghi2014}.  

By examining the distribution of kinetic energy over the
length scales, the velocity correlation function and the temporal
decay of the vortex line density, 
we show that the turbulence resulting from the thermal quench
of a Bose gas is very different from classical turbulence and
any quasi-classical regimes of superfluid turbulence. Instead, it
shares key properties 
with a second state of turbulence 
\cite{Walmsley2008,Zmeev2015} which has also been observed in superfluid
helium, called `ultra-quantum turbulence', which has unusual non-classical properties.

\section{Classical Field Method}

We model a weakly-interacting, homogeneous Bose gas
(including thermal excitations) by means of the classical field 
method \cite{Svis1,Svis2,Svis3,Svis4,Svis5,Davis,Berloff2002,Davis2,
Connaughton2005}.  
The gas is parametrized by a classical matter field $\psi({\bf r},t)$, 
normalized to the total number of particles $N=\int |\psi|^2~{\rm d}V$, 
and whose evolution follows the Gross-Pitaevskii equation (GPE)
\begin{equation}
i  \hbar \frac{\pa\psi }{\pa t}=
-\frac{\hbar^2}{2m}\nabla^{2}\psi 
+g\left|\psi \right|^{2}\psi.
\label{gpe}
\end{equation}

\noindent
The cubic term accounts for the repulsive interactions between the particles, 
with $g=4\pi \hbar^2 a_s /m$, where $m$ is the particle mass and $a_s$ is 
the inter-particle {\it s}-wave scattering length (here $a_s>0$). The total energy of 
the gas is
\begin{equation}
H=\int \left(
\frac{\hbar^2}{2m}|\nabla \psi|^2 + \frac{g}{2}|\psi|^4 \right)~{\rm d}V.
\label{eq:energy}
\end{equation}

\noindent
The GPE is conventionally used to model a zero temperature condensate, but 
it is now established that, provided the modes of the gas are highly occupied, 
the gas evolves as an ensemble of modes, each of which follows 
(to leading order) the classical trajectory described by the 
GPE \cite{Pol_Rev,Proukakis,Blakie}. Various phenomena have been studied within 
this classical field formalism, including equilibration dynamics 
\cite{Berloff2002,Davis,Connaughton2005,nazarenko_2014}, 
critical temperatures \cite{Davis2006}, 
correlation functions \cite{Wright2011}, 
vortex nucleation \cite{Sinatra2001,Leadbeater2003,Simula,Stagg2016} 
and decay of vortex rings \cite{berloff_2007}, as well as extensions to 
binary condensates \cite{Berloff_2006,Salman20091482,pattinson_2014}.

A classical field $\psi({\bf r},t)$ representing $N$ particles is simulated in a cubic periodic box of 
volume $D^3$.  We make the GPE dimensionless using the natural units of the
homogeneous system: particle number density is expressed in terms of the average value
$\rho=\braket{|\psi|^2} = N/D^3$, length in terms of the healing length 
$\xi=\hbar/\sqrt{m g \rho}$, speed in terms of the speed of sound 
$c=\sqrt{\rho g/m}$, energy in terms of the chemical potential $\mu=\rho g$, 
and time in terms of $\tau=\hbar / g \rho$.  The time evolution is computed using a fourth-order Runge-Kutta scheme with time step $\Delta t=0.01 \tau$ on a $192^3$ grid with isotropic grid spacing 
$d=0.75\xi$. 

To ensure that all numerically accessible modes 
are highly occupied, the initial condition is the highly non-equilibrium state,
\begin{equation}
    \psi\left(\mathbf{r},0\right)=\sum_{\mathbf{k}}a_{\mathbf k}\exp(i\mathbf{k}\cdot\mathbf{r}),
\label{eq:rand2}
\end{equation}
where the $192^3$ modes of the system are labelled by the wavevector ${\bf k}$, 
the coefficients $a_{\mathbf k}$ are uniform and the phases are 
distributed randomly~\cite{Berloff2002}. The occupation of mode ${\bf k}$ is
$n_{\mathbf{k}}=|a_{\mathbf{k}}|^2$.  The spatial grid discretization 
implies that high momenta are not described; in effect an 
ultraviolet cutoff is introduced, $n_{\mathbf{k}}(t)=0$ 
for $k>k_{\rm{max}}$, where $k=|{\bf k}|$ and the maximum described 
wave vector amplitude is $k_{\rm{max}} = \sqrt{3} \pi / d$ \cite{kmax}.

The condensate fraction $\rho_0/\rho$ of the equilibrium state 
(where $\rho_0$ is the density of the quasi-condensate, to be defined in Section \ref{sec:quasicondensate}) 
is uniquely determined by the number density, $N/D^3$, and the
energy density,  $\langle H \rangle/D^3$.  In practice these values
are controlled by a rescaling of the initial condition $\psi(\mathbf{r},0)$ so as to modify the uniform value of $n_{\mathbf{k}}$ in the initial condition, with a selection of high momenta coefficients $a_{\mathbf k}$ set to zero in order to retain the same number density after rescaling. The process can be seen for the initial condition in Figure \ref{fig2}.

Our results focus on the three cases: 
low condensate fraction ($\rho_0/\rho=0.22$), 
moderate condensate fraction ($\rho_0/\rho=0.48$) 
and high condensate fraction ($\rho_0/\rho=0.77$).  
The parameters used to generate these cases are listed in 
Table \ref{tbl:cond_frac}, along with the corresponding temperature of each state, $T/T_c$, where $T_c$ is the critical temperature for the condensation,  estimated using the empirical formula provided in Ref.  \cite{berloff_2007}.

\begin{table}[!ht]
\begin{ruledtabular}
\centering
\begin{tabular}{rccc}
\multicolumn{4}{c}{\it Initial conditions} \\
$N/D^3~(\xi^{-3})$           & 0.50 & 0.50 & 0.50 \\
$\langle H \rangle/D^3~(\mu \xi^{-3})$  & 2.13 & 1.33 & 0.53 \\
\multicolumn{4}{c}{\it Equilibrium state} \\
$\rho_0/\rho$        & 0.22 & 0.48 & 0.77 \\
$T/T_c$        & 0.81 & 0.56 & 0.26 \\
\end{tabular}
\end{ruledtabular}
\caption{The initial condition parameters and the resulting condensate 
fraction $\rho_0/\rho$ of the equilibrium state of the Bose gas. The quoted condensate fractions are calculated based on the choice of cutoff wavenumber $k_c=k_{c1}$. Also shown is the temperature of the gas, $T/T_c$,
estimated using the empirical formula in Ref. \cite{berloff_2007}.}
\label{tbl:cond_frac}
\end{table}

\section{Formation of the turbulent vortex tangle}

\subsection{The quasi-condensate}
\label{sec:quasicondensate}
\begin{figure}
\hspace*{-0.5cm}\includegraphics[width=0.95\linewidth]{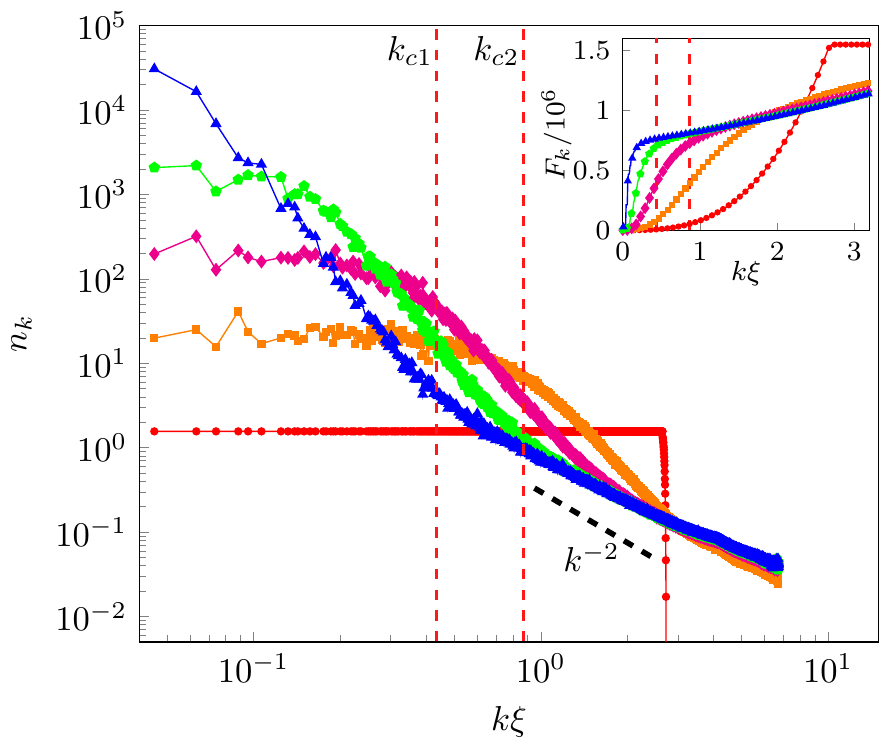}
\includegraphics[width=\linewidth]{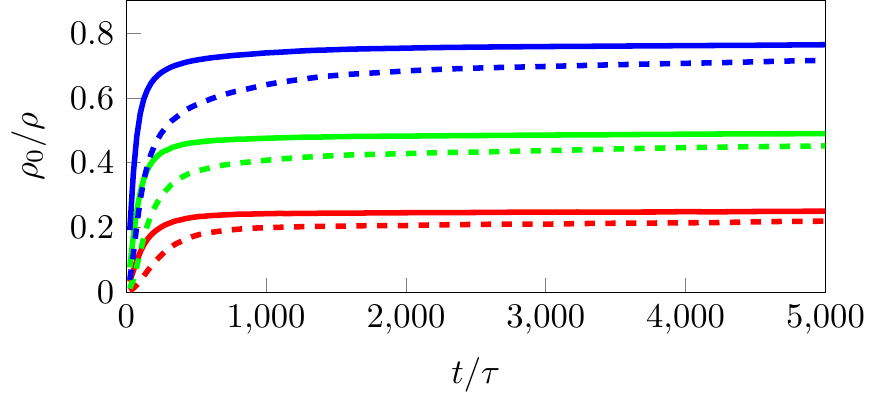}
\caption{(Color online). Upper: Occupation numbers $n_k$ as a function of wavenumber $k \xi$ at different times $t/\tau$ as the Bose gas decays to equilibrium:
$t/\tau=0$ (red circles), 
$t/\tau=25$ (orange squares), 
$t/\tau=75$ (magenta diamonds), 
$t/\tau=200$ (green pentagons), 
and $t/\tau=1000$ (blue triangles). 
Here the final condensate fraction at equilibrium is $\rho_0/\rho = 0.48$. A line proportional to the Rayleigh-Jeans distribution, $n_k \propto k^{-2}$, is shown by a black dashed line as a guide
to the eye.  
Inset: Integral distribution of the particles, $F_k = \sum_{k'<k} n_{\bf k'}$ 
vs $k \xi$ at the same times as in the main figure. In both the main figure 
and the inset $k_{c1} = 10~(2\pi/D)$ and $k_{c2} = 20~(2\pi/D)$ 
are labelled by red dashed lines. Lower: Condensate fraction over time for the quasi-condensate filtered with a cutoff of $k_{c}=10~(2\pi/D)$ (dashed lines) and $k_{c}=20~(2\pi/D)$ (solid lines), with the same equilibrium states as above. We find our choices of $k_c$ lead to only minimal variation of $\rho_0/\rho$ at equilibrium.}
\label{fig2}
\end{figure}

The evolution of the system in momentum space \cite{Berloff2002,pattinson_2014} is shown in  in Fig.~\ref{fig2} (upper). At $t=0$ (red line), the mode occupation numbers 
$n_k$ are distributed uniformly with $k$, with values of the order of unity, up to the cutoff, as per the imposed initial condition.  Over time, self-ordering of the classical field leads to a smoothing of the mode occupation distribution towards a characteristic bimodal form.  At low $k$ the modes have macroscopic occupation ($n_k\gg 1$), characteristic of Bose-Einstein condensation and of the presence of the quasi-condensate.  At high $k$ the modes have low occupation, and are the thermal excitations. Their distribution approximately follows the Rayleigh-Jeans equilibrium distribution for non-interacting particles, $n_k \sim k^{-2}$ \cite{RJ}. The transition in $k$-space between these two components is most evident when viewing the  the integral distribution of the particles over the wavenumbers, $F_k = \sum_{k'<k} n_{\bf k'}$ (inset of Fig.~\ref{fig2}); here a prominent `shoulder' in the curve marks the transition from the quasi-condensate to the thermal modes. We define this transition point between the quasi-condensate and thermal gas as $k_c$, and focus on the nominal value $k_{c1} = 10~(2\pi/D)$. However, to assess the effect of the cutoff on our findings we also consider a second cutoff,  $k_{c2} = 20~(2\pi/D)$.

Due to the randomized phases in the initial condition, the system is full
of phase defects, and a dense vortex tangle 
forms in the quasi-condensate, as seen in 
Fig.~\ref{fig1} (a).
The raw wavefunction $\psi$ is too noisy to directly visualise 
the superfluid vortex tangle. Following \cite{Berloff2002}, this problem 
is overcome by defining a quasi-condensate wavefunction 
$\hat{\psi}$ via \mbox{$\hat{a}_{{\bf k}} = a_{{\bf k}}\times\max\{1-k^{2}/k_c^2,0\}$}. This procedure, which filters high-frequency modes from $\psi$,
is analogous to spatial course-grained averaging, 
so that $\hat{\psi}$ contains only the long-wavelength component 
of the classical field $\psi$.  The quasi-condensate density is then $|\hat{\psi}|^2$ and its value is $\rho_0=\braket{|\hat{\psi}|^2}$;  the condensate fraction follows as $\rho_0/\rho$, i.e. the ratio of the quasi-condensate density to the total density. The evolution of the condensate fraction is shown in Fig. ~\ref{fig2} (lower). The condensate fraction $\rho_0/\rho$ is approximately zero at $t=0$.  It immediately undergoes a rapid growth as the quasi-condensate forms.  The growth then slows and asymptotes towards its equilibrium value.  For our primary choice of cutoff, $k_{c}=k_{c1}$ the condensate fraction reaches 95\% of its asymptotic value within a time $t=600 \tau$; in comparison, the decay of the vortex tangle towards the vortex-free state typically occurs on a timescale of $t = 7000 \tau$. For the second choice of cutoff, $k_c=k_{c2}$, the growth of the condensate fraction is slightly faster and the final condensate fraction is slightly larger, but the qualitative behaviour is unchanged. We have verified that while our simulations are sensitive to the initial condition parameters shown in Table \ref{tbl:cond_frac}, the behaviours demonstrated in Fig.~\ref{fig2} are consistent over different randomized initial conditions.

The vortices are visualized, such as in Fig. \ref{fig1}, through isosurface plots of the quasi-condensate density.  The iso-surface density level varies with time according to $|\hat{\psi}|^2 = 0.05\braket{|\hat{\psi}|^2}$. This level is chosen so that the resulting vortex structures are depicted with 
approximately constant cross-sectional radius throughout the decay, and that this is similar to the radius of a vortex core at zero temperature.

\subsection{Energy spectrum}
\begin{figure}
\centering
\hspace*{-0.5cm}\includegraphics[width=\linewidth]{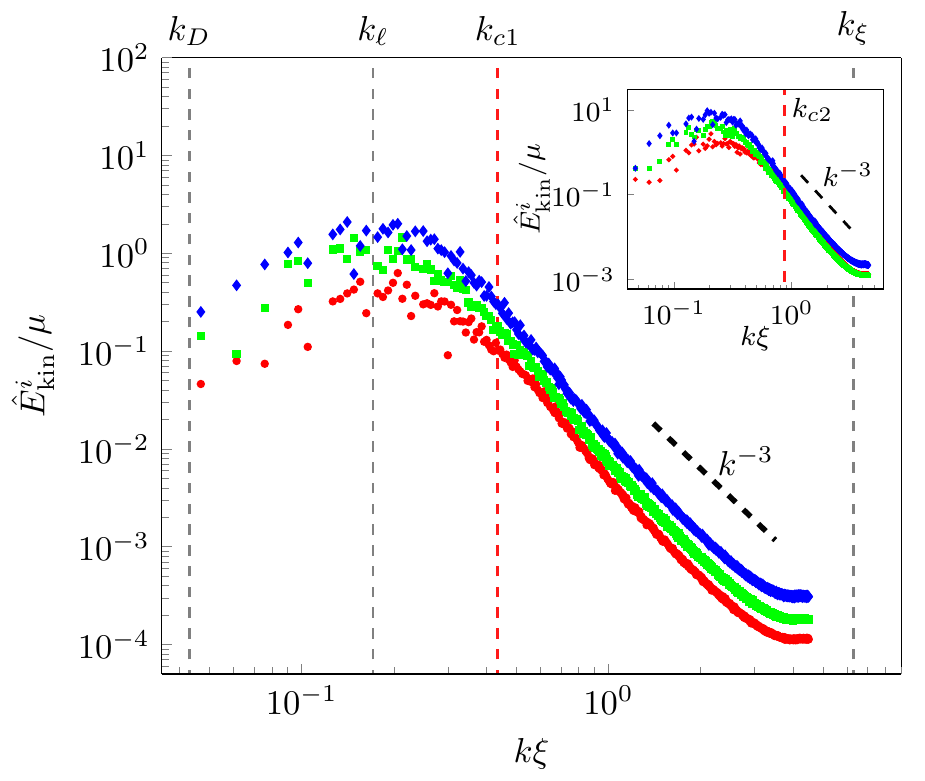}
\caption{(Color online) Incompressible kinetic energy spectrum of the quasi-condensate 
for different temperatures, corresponding
to condensate fraction $\rho_0/\rho=0.22$ 
(red circles), $\rho_0/\rho=0.48$ (green squares), 
and $\rho_0/\rho=0.77$ (blue diamonds) at time $t/\tau=300$, 
and with a cutoff of $k_{c}=10~(2\pi/D)$ (marked by a red dashed 
line labelled $k_{c1}$). 
Dashed lines also mark wavenumbers $k_D$, $k_{\ell}$ and $k_{\xi}$
corresponding to the length scale of the box, the typical inter-vortex
spacing and the healing length respectively.
A black dashed line with $k^{-3}$ dependence is plotted as a guide to the eye.
Inset: The same simulations, but with a quasi-condensate 
cutoff of $k_{c}=20~(2\pi/D)$ (marked by a red dashed line labelled $k_{c2}$).
}
\label{fig3}
\end{figure}

\noindent
One method to characterise the turbulence of the vortex tangle formed 
in the quasi-condensate is by studying how the kinetic energy of the 
vortices is distributed over wavenumber. There are three wavenumber 
scales of particular interest in this problem:
$k_D=2\pi/D$ (associated with the length scale of the computational box), 
$k_\xi = 2\pi/\xi$ (associated with the superfluid 
healing length), and $k_\ell$ (associated with the typical
inter-vortex spacing, $\ell$). The typical inter-vortex spacing can be 
estimated as $\ell \approx 1/\sqrt{L}$, where $L$ is the vortex line-density 
(length of vortex line per unit volume). These scales provide the 
perspective required to interpret the distribution of energy.

The kinetic energy density of the quasi-condensate is defined as 
$E_{\rm kin} = \frac{\hbar^2}{2m}|\nabla \hat\psi|^2$.
Using Parseval's theorem, the corresponding kinetic energy spectrum, 
$\hat{E}_{\rm kin}(k)$, can be written in terms of the angle-average 
of $\left |\mathcal{F}(\sqrt{E_{\rm kin}})\right |^2$ 
\cite{BrachetNore1997},
where $\mathcal{F}$ denotes the Fourier transform, so that,
\begin{equation}
\int E_{\rm kin}({\bf r})~{\mathrm d}V = \int \hat{E}_{\rm kin}(k)~{\mathrm d}k.
\label{eq:Ekin}
\end{equation}

\noindent
The kinetic energy can be further decomposed into compressible and incompressible parts, $E_{\rm kin} = E^i_{\rm kin} + E^c_{\rm kin}$, where the compressible part is associated with sound waves and the incompressible part with vortices. This decomposition can be defined through the hydrodynamic interpretation of the superfluid as $\sqrt{n}{\bf v} = (\sqrt{n}{\bf v})^c + (\sqrt{n}{\bf v})^i$ with $\nabla\cdot(\sqrt{n}{\bf v})^i=0$ and $\nabla \times (\sqrt{n}{\bf v})^c=0$, where $n=|\psi^2|$ is the particle density and $\mathbf{v}$ is the superfluid velocity. The incompressible kinetic 
energy spectrum is then denoted $\hat{E}^i_{\rm kin}(k)$, 
defined in a similar way to $\hat{E}_{\rm kin}(k)$ in Equation (\ref{eq:Ekin}).

 Figure~\ref{fig3} shows the incompressible 
energy spectrum of the vortex tangle at time $t/\tau=300$ (after the quasi-condensate and tangle have formed but before the 
turbulence has undergone any significant decay).  This spectrum clearly
shows that our turbulence is unlike turbulence in ordinary fluids.
In classical turbulence, the energy is concentrated
at the smallest wavenumber $k_D$, decreasing as $k^{-5/3}$ (the celebrated
Kolmogorov scaling) in the region $k_D \ll k \ll k_{\ell}$. Figure~\ref{fig3}
shows no such pile-up of energy near $k_D$ and no Kolmogorov
scaling; on the contrary, the energy peaks at length
scales just above $k_{\ell}$.  The visible $k^{-3}$ dependence of the spectra
in the region $k_\ell<k<k_\xi$ arises from the vortex cores 
\cite{BrachetNore1997}.  Note that these results are insensitive to the cutoff; we see the same behaviour for our second choice of cutoff $k_c=k_{c2}$ (inset of Fig. \ref{fig3}).

\section{Relaxation of the turbulent vortex tangle}

During the decay of the superfluid turbulence, shown in Fig.~\ref{fig1}, the vortex tangle remains 
random and isotropic 
throughout its decay to one or more vortex lines or rings, which 
eventually also decay, leading to a vortex-free state.
Insight into the nature of the turbulent decay
is obtained by monitoring the vortex line-density $L$, defined as
the vortex length
per unit volume; the vortex length is estimated as $V_{\rm t}/A$ where
$V_{\rm t}$ is the total volume within isosurface vortex tubes, and $A$
the circular cross-sectional area of a vortex, which is constant
for a steady condensate fraction.

\begin{figure}
\centering
\hspace*{-0.5cm}\includegraphics[width=\linewidth]{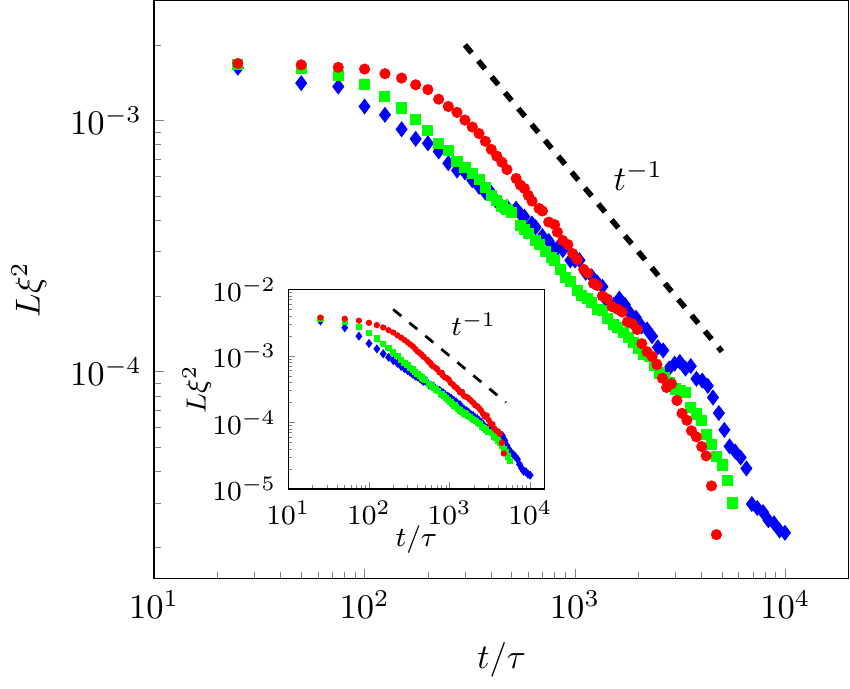}
\caption{(Color online).
Vortex line density $L$ over time $t$ for condensate fraction 
$\rho_0/\rho=0.22$ (red circles), 
$\rho_0/\rho=0.48$ (green squares), 
and $\rho_0/\rho=0.77$ (blue diamonds). 
Each line is an average of five simulations. The line proportional 
to $t^{-1}$ (characteristic of the decay of ultra-quantum turbulence)
is shown (dashed black line) as a guide to the eye.
Inset: The same simulations, but with a quasi-condensate cutoff of $k_{c}=20~(2\pi/D)$.
}
\label{fig4}
\end{figure}

\begin{table}
\begin{ruledtabular}
\begin{tabular}{cccc}
{\it $\rho_0/\rho$} & $\alpha$ &$\beta$ & {$k_c$} \\
\hline \\ [-2ex]
$0.22$ & $0.40 \pm 0.07$ & $1.04 \pm 0.03$ & $10~(2\pi/D)$\\
$0.48$ & $0.26 \pm 0.03$ & $1.03 \pm 0.02$ & $10~(2\pi/D)$\\
$0.77$ & $0.07 \pm 0.02$ & $0.83 \pm 0.05$ & $10~(2\pi/D)$\\
$0.22$ & $0.59 \pm 0.13$ & $1.04 \pm 0.04$ & $20~(2\pi/D)$\\
$0.48$ & $0.28 \pm 0.02$ & $1.05 \pm 0.02$ & $20~(2\pi/D)$\\
$0.77$ & $0.07 \pm 0.02$ & $0.84 \pm 0.03$ & $20~(2\pi/D)$\\
\end{tabular}
\end{ruledtabular}
\caption{Best fitting power law decay of the vortex line density over 
time (Figure \ref{fig4}), fitted to the equation 
$L(t) = \alpha t^{-\beta}$. The data is fitted over the region 
indicated by the length of the $L \sim t^{-1}$ guide line. The $95\%$ confidence interval for each 
fitting parameter is also indicated.}
\label{tbl:fits}
\end{table}

Figure~\ref{fig4} shows the decay of the vortex line density over time. At very early times ($t \lappeq 300 \tau$) 
the tangle is in the process of forming, while at very late times ($t \gappeq 3000 \tau$) only one or two vortices or vortex rings remain.  
However, in the large intervening range of time a sizeable tangle exists. During this range the decay of the vortex line density has a power-law form, $L =\alpha t^{-\beta t}$ with 
$\alpha$ constant and $\beta \approx 1$ (or slightly
less depending on the condensate fraction).   For comparison, the decay of classical Kolmogorov turbulence is significantly quicker, $L \sim t^{-3/2}$ \cite{Smith1993,Stalp1999,VinenNiemela2002}, and hence Fig. \ref{fig4} is consistent with Fig.~\ref{fig3} and the absence of a Kolmogorov spectrum in our system. 

Instead, the exponent $\beta \approx 1$ which we observe (see Table~II), is consistent with the 
`ultra-quantum' turbulent regime revealed by the experiments of
Walmsley and Golov \cite{Walmsley2008}, who created the turbulence by
injecting vortex rings in a sample of superfluid helium initially at rest.
In another set of experiments, Walmsley and Golov found that
a longer, more intense injection stage generates `quasi-classical'
turbulence which decays as $L \sim t^{-3/2}$. Following 
Volovik \cite{Volovik2003}, Walmsley and Golov 
argued that whereas quasi-classical
turbulence implies the existence of an energy cascade of eddies or vortex 
bundles \cite{Baggaley-Laurie-2012} similar to ordinary turbulence, ultra-quantum turbulence lacks
coherent structures and is more akin to a random flow. This interpretation
was confirmed by numerical calculations of Baggaley {\it et al.}~\cite{Baggaley2012}
who simulated the experiments, reproducing both ultra-quantum
($L \sim t^{-1}$) and quasi-classical ($L \sim t^{-3/2}$) regimes,
and verified the presence of a Kolmogorov energy spectrum
($\hat{E}_{\rm kin}(k) \sim k^{-5/3}$) only in the latter.

From the fitting parameter $\alpha$, assuming that the energy dissipation
rate per unit mass has the classical form ${\rm d}E/{\rm d}t=-\nu \omega^2$, where 
the energy per unit mass, $E$, is proportional to the length of vortex line, 
and the vorticity is estimated as
$\omega \approx \kappa L$, we infer that the turbulent kinematic viscosity
$\nu$ is such that 
 $0.06< \nu/\kappa <0.3$, in fair
agreement with numerical ($\nu/\kappa=0.06$ and $0.1$ 
\cite{Tsubota2000,Baggaley2012})
and experimental ($\nu/\kappa=0.1$ \cite{Walmsley2014})
values for superfluid helium,
although our values do not capture the temperature 
trend in helium.

\begin{figure}
\includegraphics[width=\linewidth]{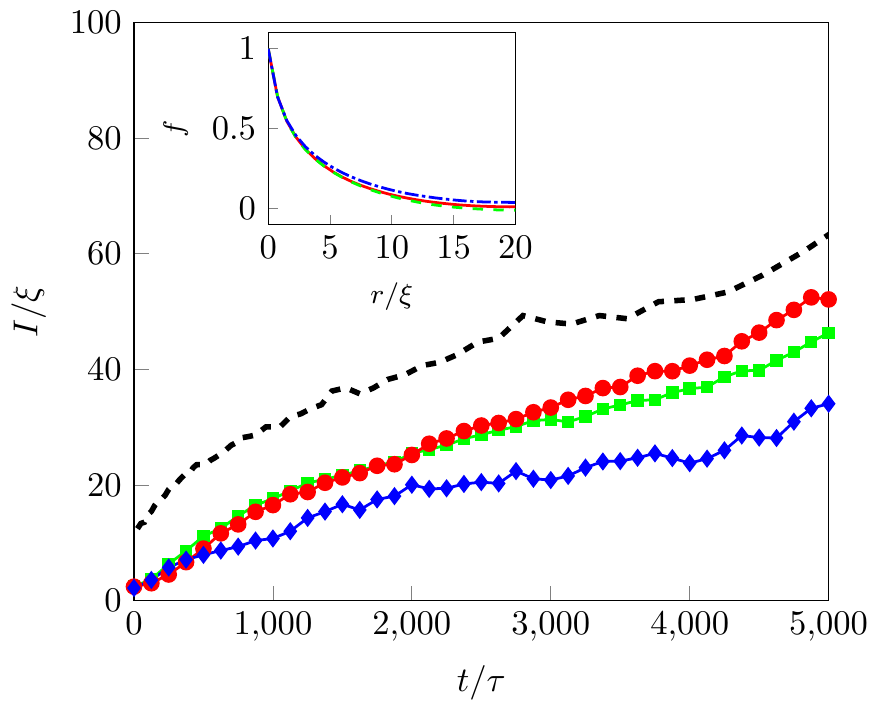}
\caption{(Color online) Integral scale $I$ as a function of time $t$ as the vortex tangle decays, 
for $\rho/\rho_0=0.22$ (red circles), 
$\rho/\rho_0=0.48$ (green squares), 
$\rho/\rho_0=0.77$ (blue diamonds),
and half the typical inter-vortex spacing $\ell/2$ (dashed line). 
Inset: Velocity correlation function $f(r,t)$ at time $t = 300\tau$ 
for equilibrium states at 
$\rho/\rho_0=0.22$ (red solid line), 
$\rho/\rho_0=0.48$ (green dashed line), 
and $\rho/\rho_0=0.77$ (blue dot-dashed line).
}
\label{fig:IS_t}
\end{figure}

To confirm the ultra-quantum interpretation of the turbulence created
by a thermal quench we compute the 
velocity correlation function, $f(r)$, defined (at fixed time $t$) as
\begin{equation}
f(r,t) = \frac{\langle v_x({\bf r},t)v_x({\bf r} +  r\hat{\bf e}_x,t) 
\rangle}{\langle v_x({\bf r},t)^2 \rangle}
\end{equation}

\noindent
where $\mathbf{v}$ is the velocity of the quasi-condensate
and the ensemble average is performed over positions ${\bf r}$.  Since the flow is isotropic, we only present results for displacements
along one direction, here chosen to be the $x$-direction. 
The function $f(r)$ is dimensionless and normalised
to $f(0)=1$. If the vortex lines are essentially randomly oriented then
at a distance $r \approx \ell/2$ the velocity correlation should vanish.
Indeed, the inset of
Fig.~\ref{fig:IS_t} shows that at time $t/\tau=300$ (when we monitor the
energy spectrum, and $\ell \approx 30 \xi$) 
the correlation function has become negligible
at such distances. In fluid dynamics,
a convenient measure of the distance over which velocities are correlated.
is the integral length scale \cite{Davidson2004}, defined  (at time $t$) as

\begin{equation}
I(t) = \int\limits_0^\infty\! f(r,t)\, \mathrm{d}r.
\end{equation}

\noindent
Figure~\ref{fig:IS_t} shows $I$ as a function of $t$
during the evolution of the vortex tangle. 
Clearly $I(0) \approx 0$ at very early times, reflecting
the random nature of the initial condition. At later times, 
$I(t)$ increases for all condensate fractions: as the tangle decays, fewer
and fewer vortices are left, and the velocity field
becomes correlated over larger regions of space. The fact that $I(t)$ remains
less than $\ell$ at all confirms the disorganized nature of our turbulence.

We find that, as with the incompressible kinetic energy and decay of vortex line-density, the qualitative behaviour of $f(r)$ and $I$ is largely unchanged by the choice of cutoff $k_c$. Filtering less modes only has the effect of introducing disorder into the quasi-condensate wavefunction, slightly reducing the quantitative values of $f(r)$ and $I$. 

\section{Conclusions}

Using classical field simulations, we have modelled the evolution of
a finite-temperature homogeneous Bose gas from a nonequilibrium
initial condition, through the formation of a turbulent tangle of vortex
lines to the relaxation to a vortex-free state.  
By monitoring the vortex line density, the energy spectrum and the
velocity correlation function, we have determined that the superfluid
turbulence created by this thermal quench process lacks the  
coherent structures and the Kolmogorov
cascade process which are typical of ordinary
(classical turbulence), and which have also been observed in
superfluid helium when driven by
propellers or towed grids, or flowing at high velocity along channels.
Instead, thermally quenched turbulence has properties similar to
another regime of superfluid turbulence called ultra-quantum turbulence,
observed in superfluid helium under certain driving conditions, which
is less organized and more similar to a random flow.

While our work is based the idealized paradigm of an infinite homogeneous system, such a system can be approximated experimentally through the recent advent of quasi-homogeneous Bose-Einstein condensates formed in box-liked traps \cite{Gaunt2013,Chomaz2015}.   Nonetheless, condensates are most commonly confined in harmonic potentials, and it would be interesting in future to explore how the inhomogeneity affects the nature of the
turbulence induced by the thermal quench.

Data supporting this publication is openly available under an Open Data Commons Open Database License \cite{data}.

\begin{acknowledgments}
G. W. S acknowledges support from the Engineering and Physical Sciences Research Council, and N. G. P. acknowledges funding from the Engineering and Physical Sciences Research Council (Grant No. EP/M005127/1). 
\end{acknowledgments}


\begin{thebibliography}{10}
\bibitem{kmax}This can be understood by considering the largest possible vector in a three-dimensional box, where each dimension of the box is of length $k_{\rm nyq} = f_s/2 = \pi/d$, the Nyquist frequency in 1D. The resulting vector is of size $k_{\rm max} = \sqrt{3}\pi/d$.
\bibitem{data} DOI to be provided on acceptance.
\end{thebibliography}
\end{document}